\documentclass{sig-alternate-10pt}
\usepackage[utf8]{inputenc}
\usepackage[english]{babel}

\usepackage{mathptmx}

\newenvironment{definition}[1][Definition]{\begin{trivlist}
\item[\hskip \labelsep {\bfseries #1}]}{\end{trivlist}}
\title{xDGP: A Dynamic Graph Processing System with Adaptive Partitioning}

\numberofauthors{4}
\author{
\alignauthor Luis M. Vaquero\\
           \affaddr{Queen Mary University of London} \\
           \email{luis.vaquero@ieee.org}
\alignauthor F\'{e}lix Cuadrado\\
           \affaddr{Queen Mary University of London}\\
          \email{felix@eecs.qmul.ac.uk}
\alignauthor Dionysios Logothetis\\
           \affaddr{Telefonica Research}\\
           \email{dl@tid.es}
\and
\alignauthor Claudio Martella\\
           \affaddr{VU University Amsterdam}\\
           \email{c.martella@vu.nl}
}
      
\begin{document}

\maketitle

\begin{abstract}
Many real-world systems, such as social networks, rely on mining efficiently large graphs, with hundreds of millions of vertices and edges. This volume of information requires partitioning the graph across multiple nodes in a distributed system. This has a deep effect on performance, as traversing edges cut between partitions incurs a significant performance penalty due to the cost of communication. Thus, several systems in the literature have attempted to improve computational performance by enhancing graph partitioning, but they do not support another characteristic of real-world graphs: graphs are inherently dynamic, their topology evolves continuously, and subsequently the optimum partitioning also changes over time.

In this work, we present the first system that dynamically repartitions massive graphs to adapt to structural changes. The system optimises graph partitioning to prevent performance degradation without using data replication. The system adopts an iterative vertex migration algorithm that relies on local information only, making complex coordination unnecessary. We show how the improvement in graph partitioning reduces execution time by over $50\%$, while adapting the partitioning to a large number of changes to the graph in three real-world scenarios.

\end{abstract}



\section{Introduction}

Many critical analytical tasks in real-world
systems, such as ranking and recommending online content or discovering groups of
correlated stocks, depend on the ability to mine graphs
of hundreds of millions of vertices and billions of edges. The importance of managing
graphs of that scale is evidenced by the emergence of  numerous distributed
graph storage and graph processing systems in recent years~\cite{
Cheng2012, Power:2010:PBF:1924943.1924964, giraph, trinity,
Seo:2010:HEM:1931470.1931872, low2010, Krepska2011, Stutz2010}.

Most graph processing systems adopt a batch job model, greatly
influenced by  Pregel  \cite{Malewicz2009}: A static graph is loaded in the memory of a distributed system to cope with its very large scale. Computations are performed in multiple iterations that involve per-vertex processing and messaging exchange between neighbouring vertices. Graph partitioning is crucial to performance in these systems: Balanced graph partitioning helps with load balancing, and minimising the number of cuts between partitions improves communication cost between neighbouring vertices~\cite{Malewicz2009}. 

These systems process static graphs, but many real-world graphs are dynamic (the graph structure changes over time) and several 
analytical applications require near real-time response to graph changes. For instance, the Twitter graph may receive thousands of updates per second at
peak rate~\cite{twitterrate} that can potentially indicate new trending topics.
Topic recommendation systems must reflect these changes within minutes,
otherwise they become irrelevant.  Similarly, telecommunications operators must detect fraud by
mining the Call Detail Record (CDR) graph in real-time \cite{Weigert2011}. In addition to
the existing batch processing systems, there is a need for frameworks that simplify
continuous processing of dynamic massive graphs.

Supporting  dynamic graphs brings new challenges. As the graph structure changes over time, if partitions were not updated, performance would constantly be degraded due to additional communication overhead and unbalanced partitions (processing bottlenecks). However, these are often conflicting requirements that make optimisations difficult and more so when rapid responses are needed. 
 
In this paper, we present a system for processing large-scale dynamic graphs. We 
have addressed the performance challenges described above by implementing a
scalable partitioning heuristic with minimum overhead. Our heuristic is based on decentralised, iterative
vertex migration. The heuristic migrates vertices between partitions trying to minimise the number of cut
edges, while at the same time keeping partitions balanced upon structural changes at run time. Updates in the graph topology trigger the iterative vertex migration process that adapts partitioning to the new topology.

This paper presents the following contributions:
\begin{enumerate}
\item We report how changes in the structure of a graph impact the quality of the graph partitioning, as even high quality initial partitioning strategies failing to adapt to graph dynamics.
\item We describe a system that processes continuously large-scale dynamic graphs, automatically adapting to structural changes on the graph for improved performance.
\item We present a highly scalable, completely decentralised heuristic based on label propagation.
 The heuristic improves graph partitioning in the event of dynamic changes in the topology of the graph relying only on local information.
\item We present extensive evaluation of the heuristic through
system deployments, using synthetic and real datasets. Experiments show the effectiveness of the heuristic in adapting to graph changes, the associated performance improvement, and the observed impact in different real scenarios. 

\end{enumerate}

The rest of the paper is organised as follows. Section~\ref{sec:motiv} presents 
the motivation for addressing partitioning of dynamic graphs. In Section~\ref{sec:aip}, we
describe our heuristic in more detail. Section~\ref{sec:lsid} describes some
relevant pitfalls that need to be overcome when realising the heuristic into a
real system. The heuristic is then tested at scale in a series of lab
experiments and real-world use cases in Section~\ref{sec:eval}.
Section~\ref{sec:related} compares our contributions to the related
work. We present the main conclusions and discussion in
Section~\ref{sec:conclusions}.

\section{Dynamism in Graph Partitioning}
\label{sec:motiv}

The number of cut edges in a distributed graph processing system directly impacts the communications overhead of computations across the whole graph, up to the point of becoming a key performance factor~\cite{Malewicz2009}. The performance impact of graph partitioning has led to several optimisations at the beginning of the processing, right when the graph is being loaded in memory. For instance, popular heuristics for content ranking converge faster if initialised with smarter graph partitioning heuristics~\cite{Tong2006}. At load time, clever partitioning heuristics to improve performance in massive graphs have also been employed \cite{Ugander,Salihoglu2012,Stanton2012}. None of them is capable of preventing performance degradation arising from changes to the graph structure over time.

\begin{figure}
  \centering
    \includegraphics[width=0.98\columnwidth]{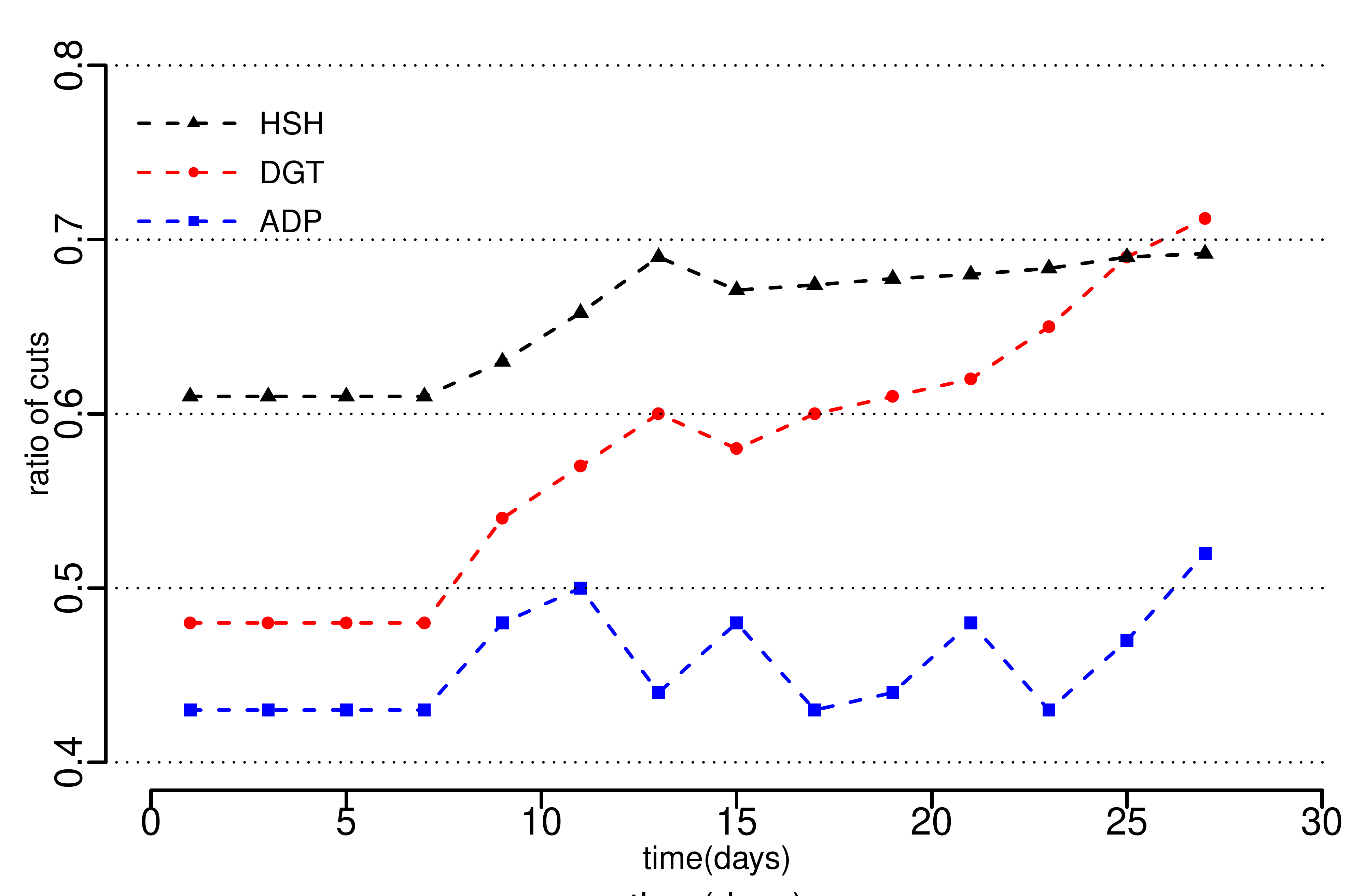}
       \vspace{-10pt}
  \caption{ Evolution of the ratio of cuts over time on a dynamic graph generated by processing CDR data on a call
  window. }
         \vspace{-10pt}
  \label{fig:2-dyngraphs}
\end{figure}

To illustrate the impact of dynamism in graph partitioning, we built a call graph from mobile CDR data (more details are shown in Section \ref{sec:eval}). The graph was partitioned using three different techniques: \emph{modulo hash} (HSH), the most popular technique because of its high scalability to produce balanced partitions, results in high communication overheads \cite{Mizan}; a state of art streaming partition technique (\emph{deterministic greedy}, DTG) \cite{Stanton2012}; and our \emph{adaptive repartitioning} heuristic, as described below, (ADP). Figure \ref{fig:2-dyngraphs} shows  the evolution of the partitioning (expressed in the percentage of edges that cut across different partitions). As the graph evolves over time, a static or streaming partitioning solution cannot cope with the changes and, consequently,
the quality of the partitioning (and the system performance) degrades over time. 

Current approaches either allow partitioning to degrade as the graph changes or require graph re-partitioning, which is a very costly process on large-scale graphs. Both of them effectively increase processing time. While this problem does not deeply affect batch processing systems, it can greatly impact the throughput and latency of graph processing systems requiring faster response times.

To enable graph mining applications in real-world environments,
such as online social networks, we need scalable partitioning heuristics that
take the dynamic nature of the graphs into consideration. This
requires the ability to quickly adapt the partitioning as the
graph changes to prevent performance degradation.

\section{Adaptive Iterative Partitioning}
\label{sec:aip}

In this section the iterative heuristic for improved dynamic graph partitioning is presented. Before we describe the heuristic, the next subsection describes the problem and context that any solution should address.

\subsection{Problem Statement}

\begin{definition} 

\emph{(Dynamic Graph Partitioning)}
A dynamic graph $G(t) = (V(t), E(t) )$ is a graph whose vertices $V$ and edges $E$ can change over time $t$, either with addition or removal of elements. Let $P(t)$ be the set of partitions on $V$ at time $t$ and  $P^{i}(t)$ the individual partition $i$, with $|P^{i}| = k$. These partitions will be such that $\bigcup\limits_{i,t}^{k} P^{i}(t) = V$ and $P^{i}(t) \cap P^{j}(t) = \emptyset$ for any $i \neq j$. The edge cut set $E_{c}\subseteq E$ is the set of edges which endpoint vertices belong to different partitions.
\end{definition}

A distributed graph processing system splits the partitions between nodes. Every vertex belonging to the graph has an assigned partition.  At time $t = 0$, the graph is loaded  with an initial partitioning.  New vertices appearing at $t > 0$ also have to be assigned to a partition according to a strategy.The most commonly used strategy in large-scale graph processing systems is \emph{hash partitioning}. Given a hashing function $H(v)$, a vertex is assigned to partition $P^{i}(0)$ if $H(v)$~\emph{mod}~$k = i$. This strategy is effective as it is lightweight, it does not require a global lookup table, and, depending on the characteristics of the vertex IDs, it can scatter the vertices uniformly across the partitions. Unfortunately, it introduces many cut edges. In addition, this method does not guarantee adaptation to changes in the topology of the graph, since its initial partitioning is never updated. 

The goal of repartitioning is to minimise the number of cut edges across partitions, while keeping the partitions balanced in order to improve application performance. Reducing the number of cut edges diminishes the communications overhead of computations across the whole graph, while load balancing can have a significant impact on overall processing time~\cite{Malewicz2009}. The characteristics of dynamic graphs and the context of application also bring several challenges that must be considered for any solution:

\begin{itemize}

\item \emph{Graph structure changes are not predictable}. It is not possible to know beforehand the nature of the changes, or the impact they will have in performance, unless working on a specific use case. Therefore, partition optimisation should apply to any change to the graph, and must be able to run continuously if the graph changes all the time.
\item \emph{The computational overhead from the partitioning optimisation must be low}. As the objective is to improve application performance, the selected technique must be lightweight and scalable to work over large graphs.
\item \emph{Synchronising distributed state at a large scale is very costly in a dynamic environment}. Computations will have a partitioned view of the complete system state. Propagating global information across the network incurs a significant overhead, which must be considered for every repartitioning technique. 

\end{itemize}

\subsection{Greedy Vertex Migration}
\label{subsec:greedy}

We have defined a heuristic for dynamically adapting graph partitions that considers the challenges above. The heuristic is based on label propagation \cite{raghavan2007}, adapting the approach of performing iterative computations based on per-vertex neighbour information. Label propagation was first proposed as an efficient method for learning missing categories in semi-supervised learning scenarios. Unlabelled nodes iteratively adopt the label of the majority of their neighbours until no new labels are assigned (convergence is reached). The technique has been adopted in the literature for supporting adaptive migrations on static graphs\cite{Ugander}. 

The iterative heuristic works as follows.  On every iteration $t$\footnote{Note that we measure time in number of iterations, decoupling the heuristic from implementation considerations. The actual time taken by an iteration to complete will depend on the system and the specific load of the system at that iteration.} after the initial partitioning, each vertex will make a decision to either remain in the current partition, or to migrate to a different one. Migration decisions are only based on local information available to the vertex, where the goal is to ``get neighbours together'' in order to minimise the number of cut edges $|E_c|$. At the end of the iteration, all vertices who decided to migrate will change to their desired partitions. Video 1\footnote{https://dl.dropbox.com/u/5262310/reducedCuts.avi} shows how the heuristic evolves evolves partitioning over time in a 2d slice of a 3d cube of a $10^6$ vertices mesh graph, where every vertex is physically surrounded by its neighbours. As time goes, the initial hash partitioning across 9 partitions (represented with a different colour each) is improved by increasing the number of neighbours placed together.

Dynamism comes natively in this iterative approach. New vertices are initially assigned a partition according to an strategy (we opted for hash modulo) and the heuristic will attempt automatically to move them closer to their neighbours. 

For vertex migration decisions we evaluated multiple alternatives based on local information \cite{Stanton2012,Prabha12}. We chose a greedy heuristic that had the lowest computational cost while yielding strong results regarding minimising cut edges. The heuristic works as follows: At each iteration, a vertex will decide to migrate to the partition with the highest number of neighbour vertices. With this premise, the candidate partitions for each vertex are those where the highest number of its neighbours are located. Formally, for a vertex $v$, the list of candidate partitions is derived as follows: $cand(v,t) = \{ P^{i}(t)\in P(t), \exists$ w, w $\in (P^{i}(t) \cap \Gamma(v,t)  ) \}$, where $\Gamma(v,t)$ is the set of $v$ plus its neighbours at iteration $t$. Since migrating a vertex potentially introduces an overhead, the heuristic will preferentially choose to stay in the current partition if it is one of the candidates.

The heuristic relies on local information, as each vertex $v$ only needs to know the location of its neighbours in order to choose its destination. In a real system, that information will already be available at each partition (in order for vertices to communicate with their neighbours). The heuristic does not need of additional coordination mechanisms for sharing further information, which might hamper scalability.

\subsection{Maintaining Balanced Partitions}

The greedy nature of the presented heuristic will naturally cause higher concentration of vertices in some partitions. We refer to this phenomenon, common to general label propagation algorithms \cite{raghavan2007}, as node densification. As our goal is to obtain a balanced partitioning, we set a capacity limit  for every partition.
\begin{definition} \emph{(Partition Capacity)}.
Let $C^{i}$ be the capacity constraint on each partition. At all times $t$, for each partition $i$, $|P^{i}(t)| \leq C^{i}$. 
\end{definition}

In order to control node densification, vertices need to be aware of the maximum partition capacities capacity ${C^{i}}$. The remaining capacity of each partition $i$ at iteration $t$ is $C^i(t) = C^{i} -|P^i(t)|$. These values change every iteration, forcing to relax our local information constraint.
 
The local and independent nature of migration decisions make these capacity limits difficult to enforce. At iteration $t$ the decision of a vertex to migrate can only be based on the capacities $C^i(t)$ computed at the beginning of the iteration. These capacities will not be updated during the iteration, which implies that without further restrictions all vertices will be allowed to migrate to the same destination, potentially exceeding the capacity limit. 

We ensure these limits will not be surpassed by independent  decisions by working on a worst case basis. We split the available capacity for each partition equally and we use these splits as quotas for the other partitions. Hence, the maximum number of vertices that can migrate from partition $i$ to partition $j$ over an iteration $t$ is defined as: $Q^{i,j}(t)=\frac{C^j(t)}{|P(t)|-1}$; $j \neq i$. See Section \ref{sec:lsid} for system implementation details  

Quotas can defer migration decisions, which has a side effect on the real system performance. On a real system, vertex migrations are a costly activity than can affect application performance: They involve messaging across partitions, object creation/destruction, and memory reservation, incurring on significant overhead. A lesser number of migrations per iteration reduces the extra load imposed by the migration decisions to the system, as well as the maximum overhead imposed to the system in a single iteration.

This strategy to manage partition capacity introduces minimum coordination overhead. Vertices base their decision on the location of their neighbours, and the partition-level current capacity information, which must be available locally to every node. Propagating capacity information is scalable, as it is proportional to the total number of partitions $k$.

\subsection{Ensuring Convergence} 

The independent nature of the migration decisions endangers convergence of the heuristic. Local symmetries in the graph may cause pairs (or higher cardinality sets) of neighbour vertices independently decide to ``chase each other'' in the same iteration, as the best option is to join its neighbour. 

We have addressed these issues by introducing a random factor to the migration decisions. At each iteration, each vertex will decide whether to migrate with probability $s$, $0 < s < 1$. A value of $s=0$ causes no migration whatsoever, while $s=1$ allows vertices to migrate on every iteration they attempt to. Intermediate values in the range address the chasing effect, but lower values also affect the overall convergence time.

\begin{figure}
  \centering
    \includegraphics[width=8cm]{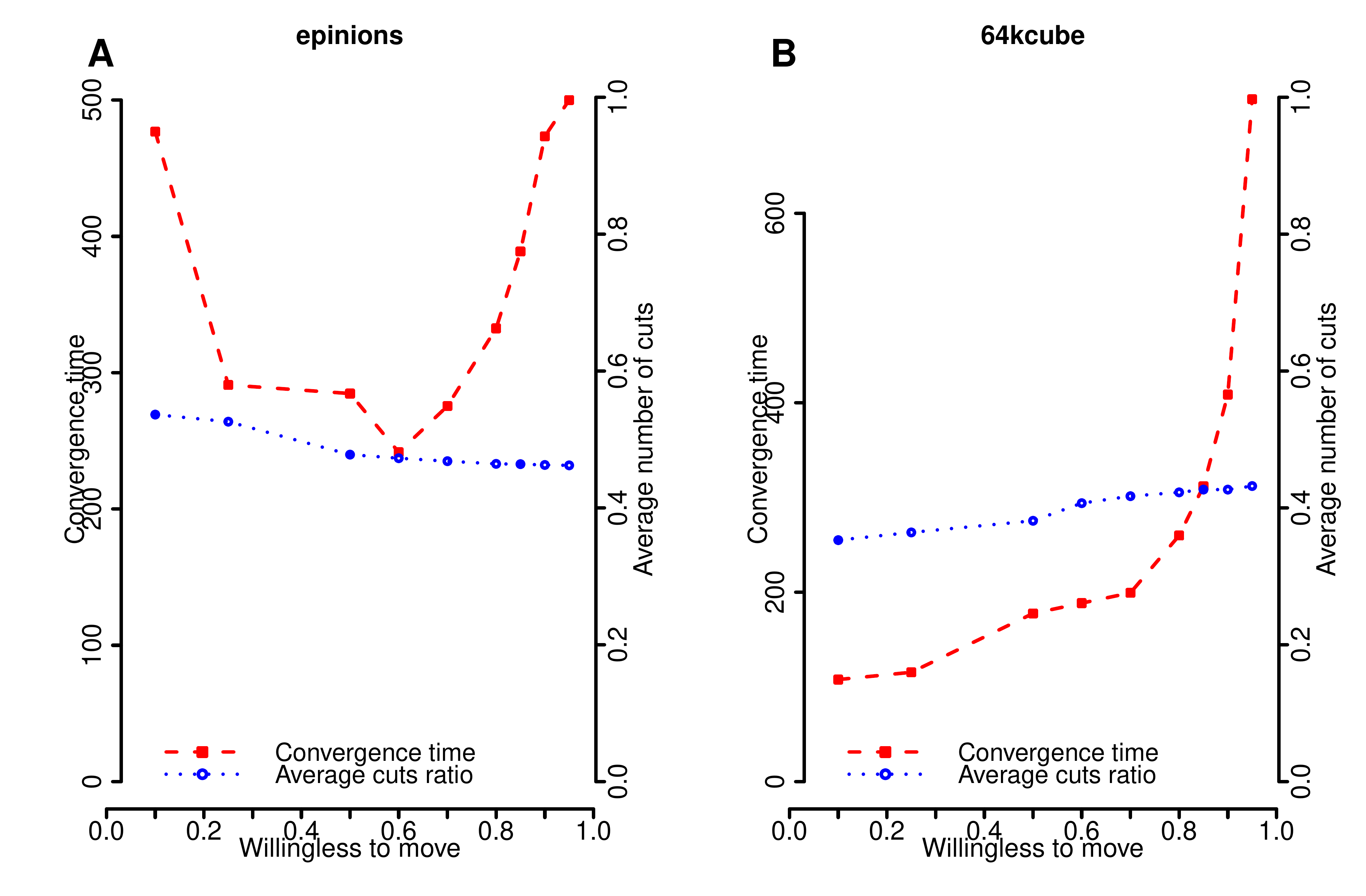}
  \caption{Effect of $s$ into Convergence and Number of Cuts (normalised to the total number of edges in the graph). Average of 10 experiments performed over two graphs: 64kcube (A) and Epinions(B) from Table \ref{table:summary}, partitioning over 9 nodes.}
    \vspace{-10pt}
  \label{fig:convergence}
\end{figure}

We explored the effect of different values of $s$ with an extensive set of experiments on different graphs, assessing convergence time and node densification. Details about the selected graphs are provided in Section \ref{sec:dataset}. We assumed full convergence when the number of vertex migrations was zero for more than 30 consecutive iterations. Figure \ref{fig:convergence} shows the effect of $s$ on convergence time and normalised number of cuts for two different graphs. In both cases, there was no statistical difference in the number of cuts achieved by the heuristic, regardless of the value of $s$. Similar results were obtained for the remaining graphs used in our study, shown in Table \ref{table:summary}.

However, $s$ can have a significant impact on convergence time. Low values of $s$ limit the number of migrations executed per iteration, potentially increasing the time required for convergence. On the other side, high values fail to fully compensate the neighbour chasing effect, introducing wasted migrations per iteration that delay convergence and increase computation time. This is particularly evident in Figure \ref{fig:convergence} (B). From our experience, a constant intermediate value ($s = 0.5$) will have adequate performance over a variety of graphs: the reduced message overhead makes processing differences (due to variations in $s$) negligible. This is specially true in the context of long running (continuous) processing systems.

\section{System Design}
\label{sec:lsid}

In this section, we present xDGP, our large-scale dynamic graph processing system. We provide an overview about the system computation model, the distributed system architecture, and finally detail how we have integrated the iterative adaptation heuristic.

\subsection{Computation Model} 

xDGP takes Pregel's computational model (``think like a vertex'') as inspiration and expands it to a continuous dynamic graph processing model, instead of focussing on batch jobs. Computation is composed of iterations, where the same function is executed for each vertex. A job defines a function applied to each vertex, synchronising messages at the end of every iteration. 

A job starts processing an initial graph, which is loaded into the system with an initial partitioning. The system also provides an external API for modifying the topology of the graph at any time. Functions allow adding and removing vertices and edges from the graph. API topology change requests are added to a change queue, and are processed at the end of every iteration (or potentially, after $n$ iterations, as shown in the CDR case (Section \ref{subsec:usecases}). 

At the start of every computing iteration, an iteration of the  adaptive migration heuristic runs over the graph, potentially triggering decisions to adapt the updated graph to the last changes to the graph.

\subsection{Vertex Migration Support}

In this subsection we provide the main insights derived from our experience implementing xDGP.

\begin{figure}
 \centering
    \includegraphics[width=5.cm]{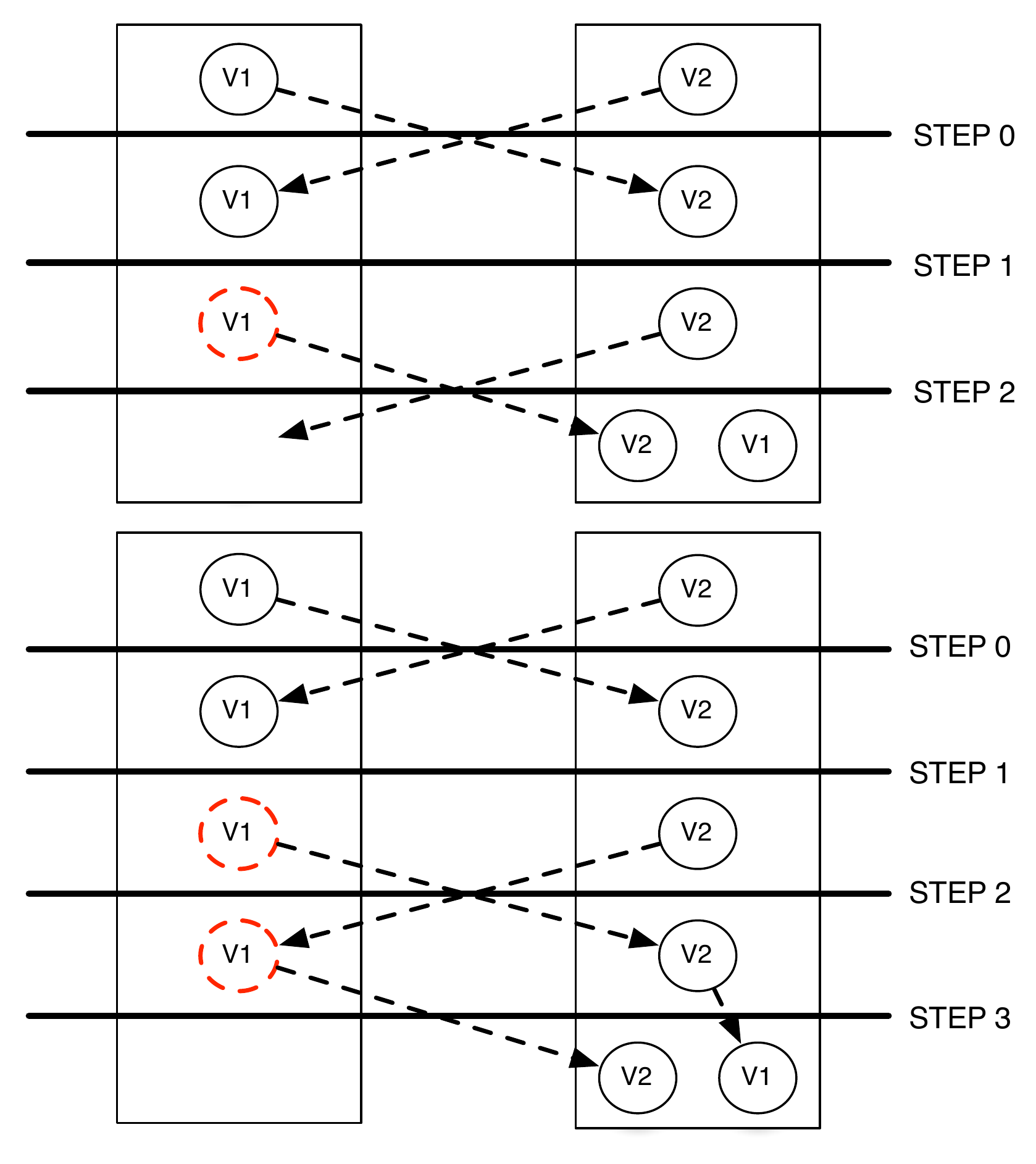}
 \caption{ Deferred Vertex Migration to Ensure Message Delivery. \textit{Top: } Failed message delivery due to incorrect synchronisation. \textit{Bottom: } Correct delivery. The dashed-red circle indicates when the vertex is in a ``migrating'' state waiting for one iteration (step) before actually migrating.}
 \label{fig:dance}
 \vspace{-10pt}
\end{figure}

\paragraph{\textbf{Deferred Vertex Migration}} At any iteration $t$, vertices take independent migration decisions, and potentially send messages to be processed by their neighbours. At $t$, a vertex does not know the destination of neighbour vertices at $t +1$ (this would require immediate notification to their neighbours of the migration decision, that would need to be received and processed in the same iteration, increasing the overhead from the migration process). Migrating a vertex at the next iteration after its decision would require one of the following adaptations to avoid losing messages (see Figure \ref{fig:dance} (top)): either forwarding the incoming messages to the new destination of the vertex, or updating the messages in the outgoing queues of the other workers with the updated destination. However, these solutions require additional synchronisation and coordination capabilities that would challenge the scalability of the  heuristic.

Instead, we solved this coordination problem with no additional overhead: We force vertices to wait for one iteration before they migrate. At the end of iteration $t$, at which the vertex requested the migration, the host  worker sends a message to the other workers about the upcoming migration, so that they will have been notified at the start of the following iteration $t +1 $, and the new messages produced during iteration $t+1$ can be sent directly to the new destination (see Figure \ref{fig:dance} (bottom)). This way the computation is not directly affected by the migrations.

\paragraph{\textbf{Worker to Worker Capacity Messaging}} The heuristic requires the system to maintain an extra element of global information: Each worker must notify the $|C^i(t)|$ of its partitions to the other workers. We implemented these communications using the scalable messaging form our computation model, with the restriction that messages will only be received and processed by the next iteration.  Therefore, when using this mechanism, workers  send information about their capacity at iteration $t+1$, ensuring partial freshness. The predicted capacity will be $C^{i}(t+1) = C^i(t) - V_{out}^{i}(t+1)+ V_{in}^{i}(t+1)  $, where $V_{in}^{i}(t) \subset V $ are the vertices migrating to $i$ in $t+1$, and $V_{out}^{i}(t) \subset V $ are the vertices migrating from $i$ to other partitions in $t+1$. $V_{out}^{i}(t+1)$ is known by the worker as it is based on local decisions.  $V_{in}^{i}(t+1)$ is also known by the worker at iteration $t+1$ as deferred vertex migration ensures that the workers will be aware of this value. 

\begin{figure*}
\centering
\includegraphics[width=12cm]{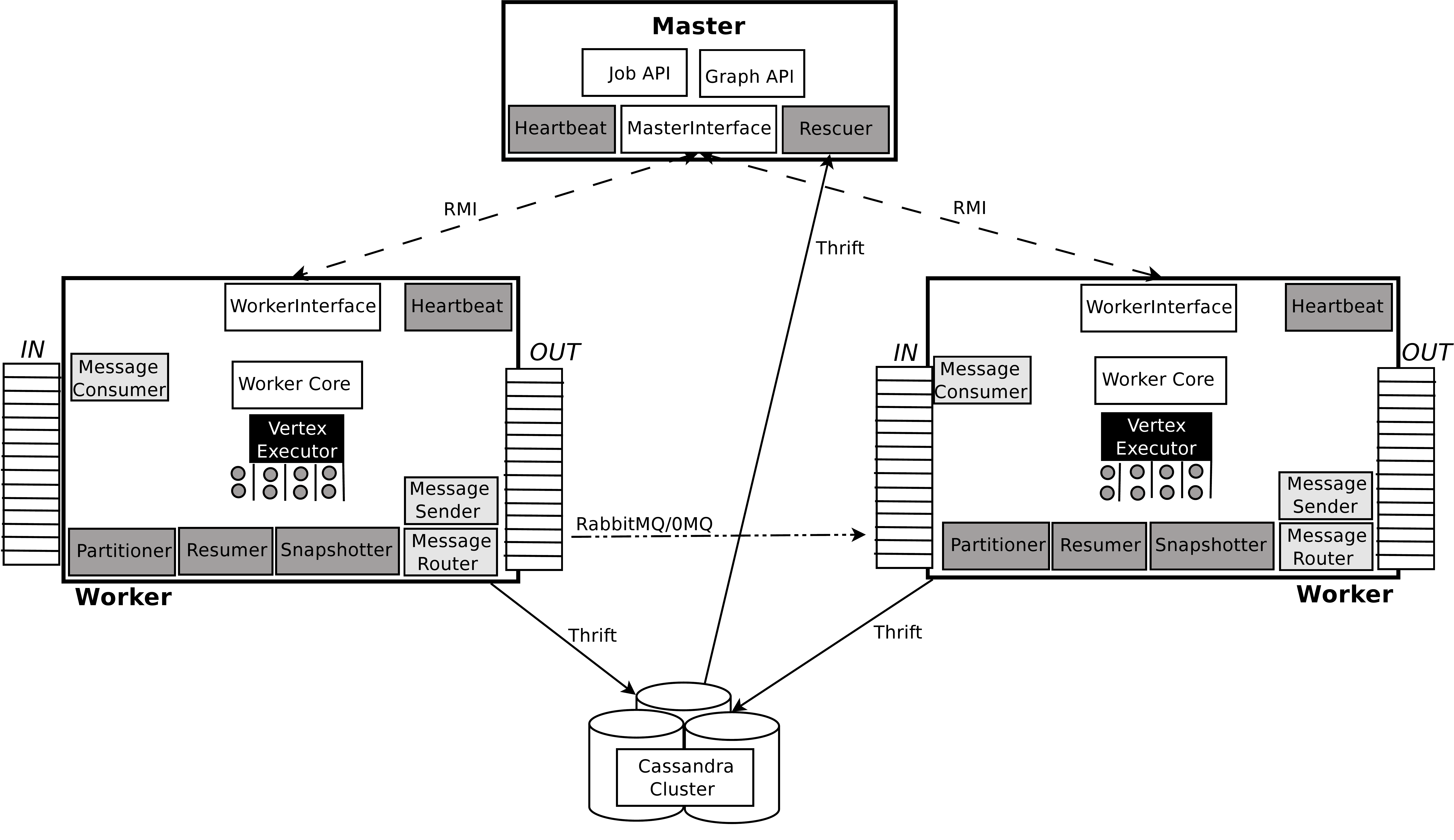}
\caption{ xDGP overview. Small grey dots represent the vertices and the vertical lines under the Vertex Executor box indicate that several concurrent executors run in parallel in a multi-core host.}
\label{fig:overview}

\end{figure*}

\subsection{System Implementation}

xDGP  is implemented in Java, and has as main design goals support for dynamic graph adaptation, failure tolerance, and intermediate results snapshot. Following the Bulk Synchronous Processing \cite{Valiant1990} model, the main system blocks are Master and Workers, as shown in Figure \ref{fig:overview}. Master and workers communicate synchronously (RMI) to enforce the global synchronisation barrier. Similarly to Pregel, xDGP implements an abstract Vertex class that hides all the complexity to the user, with the system orchestrating the vertex-level computations in parallel across the workers. An execution controller creates a number of threads, depending on the number of CPUs available. 

Workers keep input and output message queues for inter-worker vertex communications, sending messages through a loosely coupled asynchronous delivery method (we used RabbitMQ and ZeroMQ as interchangeable message handlers). xDGP also provides a generalisable message aggregation mechanism (grouping messages sharing same source and destination workers, in order to reduce the routing inefficiency introduced by RabbitMQ).

Snapshots (for failure-tolerance or for keeping the output of the running heuristic) are stored on a distributed-column store cluster (replicated configuration), keeping balance between writing speed and consistency. xDGP tries to find the right balance between failure-tolerance and high write throughput (so as not to delay computation). Frequent snapshotting and high write-throughput are specially important for dynamic graphs, since intermediate analysis results must be kept for the external applications to show the output and its evolution.

The system supports two types of messages. Application (vertex to vertex) messages contain data related to the computation, and system messages (worker to worker) support information exchange (e.g. notifying current capacity to other workers). While system messages routing is straightforward, dynamic vertex migration makes routing of application messages more complicated. A Vertex Locator in each of the workers helps to find the current location of that vertex.

When new vertices are added, the partitioner of each of the worker nodes is in charge of properly allocating them to one of the workers. This decision is important to reduce convergence time: a random placement strategy works just as well since that new vertex would be migrated around until it finds most of its neighbours are local. This approach is preferred to more complex placements (that involve more coordination messages and delay the migration process that occurs in between two iterations). Buffers (distributed queues) are in charge of dampening new requests to add/delete graph elements. Queues for vertex or edge deletion/addition can be prioritised.

\section{Evaluation}
\label{sec:eval}

In this section, we present the evaluation experiments we have undertaken for testing the dynamic graph adaptation capabilities of the proposed system and heuristic. First, we validate the capabilities of the heuristic through a set of micro-benchmarks, looking at different aspects of adaptive migration (quality of the partitioning, convergence time and distribution of migrations, and absorption of graph changes). Second, we demonstrate through three use cases how the system and heuristic can improve the processing of real scenarios, namely real-time social network analysis, online queries from CDR records, and large-scale biomedical simulations.

\subsection{Datasets}
\label{sec:dataset}

We have selected a diverse collection of graphs (see Table \ref{table:summary}), including synthetic graphs and real-world graphs, of multiple sizes (up to 300 million edges) and edge distributions: homogeneous finite-element meshes (FEM) and power-law degree distribution. 

Regarding the synthetic graphs, the synthetic meshes have a 3d regular cubic structure, modelling the electric connections between heart cells \cite{Tusscher2004}. Power law synthetic graphs have been generated with networkX \cite{networkx}, using its power law degree distribution and approximate average clustering \cite{powerlawgen}; the intended average degree is $ D = log(|V|) $, with rewiring 
probability $p =0.1$. 

When required, we mimicked dynamic changes to the synthetic graphs by adding nodes and vertices using the well-known ``forest fire'' model \cite{Leskovec2007}, and updating the graph with these additions in a single step.

\begin{table}[h!b!p!]
\vspace{-10pt}
\caption{ Summary of the evaluation datasets.}
\scriptsize
\label{table:summary} 
\center
    \begin{tabular}{ | c | c | c | c | c |}
    \hline
    Name & $|V|$ & $|E|$ & Type & Source \\ \hline
      \textit{1e4} & 10000 & 27900 & FEM & synth \\ \hline
      \textit{64kcube} & 64000 & 187200 & FEM & synth \\ \hline
      \textit{1e6} & 1000000 & 2970000 & FEM & synth \\ \hline
      \textit{1e8} & $10^8$ & $2.97 *10^8$ & FEM & synth \\ \hline
      \textit{3elt} & 4720 & 13722 & FEM & \cite{femgraphs} \\ \hline
      \textit{4elt} & 15606 & 45878 & FEM & \cite{femgraphs} \\ \hline
      \textit{plc1000} & 1000 & 9879 & pwlaw & synth \\ \hline
      \textit{plc10000} & 10000 & 129774 & pwlaw & synth \\ \hline
      \textit{plc50000} & 50000 & 1249061 & pwlaw & synth \\ \hline
      \textit{wikivote} & 7115 &  103689 & pwlaw & \cite{wikivote} \\ \hline
      \textit{epinion} & 75879 & 508837 & pwlaw & \cite{epinions} \\ \hline
       \textit{livejournal} & 4847571 & 68993773 & pwlaw & \cite{lwa} \\ \hline       
    \end{tabular}
\end{table}

In addition to these graphs, we also used two real-world sources of dynamic data: 

\begin{enumerate}
\item We processed tweets from Twitter's streaming API  in real-time for a week, generating nodes from users and edges from user mentions in tweets; 
\item We processed one-month data of anonymised calls in a mobile European operator. We fed these data chronologically, building a dynamic graph of call interactions, consisting of 21 million vertices, 132 million reciprocated social ties. 

\end{enumerate}

\subsection{Microbenchmarks}
\label{sec:ubenchmark}

The goal of these experiments is to understand the quality, performance and cost of the adaptive migration heuristic. For quality, we adopted the \emph{cut ratio}, i.e. the ratio of edges cutting across different graph partitions. From the application performance point of view, the lower the cut ratio the lower the number of communication messages that will be sent across the distributed system.

We estimate the runtime overhead of the heuristic  by characterising how the heuristic triggers \emph{migration of vertices} over multiple iterations. 

Finally, we compare our estimations with the measured performance of the heuristic (average \emph{computation time} of an iteration). 

We provide for comparison the results obtained by running the same experiments in our system, without adaptive partitioning. All the experiments shown below are the mean of $n=10$ repetitions. Variability is reported in the form of standard deviation in the error bars. \newline
 
 
\subsubsection{Quality of partitioning}

As we mentioned above, some systems just optimise the initial loading of the graph in memory, but provide no later adaptation to cope with structural changes in dynamic graphs. Different initial partitioning techniques split the graph in different ways that may affect the way our heuristic behaves during run time adaptations. 

In this experiment we used different graphs and studied if our heuristic could lead to further improvements in performance on nine partitions. We tested several initial partitioning strategies: 1) \emph{Hash Partitioning (HSH)}: the destination partition is computed by hashing vertexId with modulo number of partitions; 2) \emph{Pseudorandom Partitioning (RND)}: vertices were assigned to partitions through a pseudorandom generator, still ensuring balanced partitions; 3) \emph{Deterministic Greedy (DGR)}: stream-based ``linear deterministic greedy'' as presented in \cite{Stanton2012}; 4) \emph{Minimum Number of Neighbours (MNN)}: applies the same stream-based approach to the ``minimum number of neighbours'' heuristic presented in \cite{Prabha12}. After initialising the partitions with one of four different initial strategies, we ran our adaptive iterative heuristic until convergence. 

\begin{figure}
  \centering
    \includegraphics[width=0.98\columnwidth]{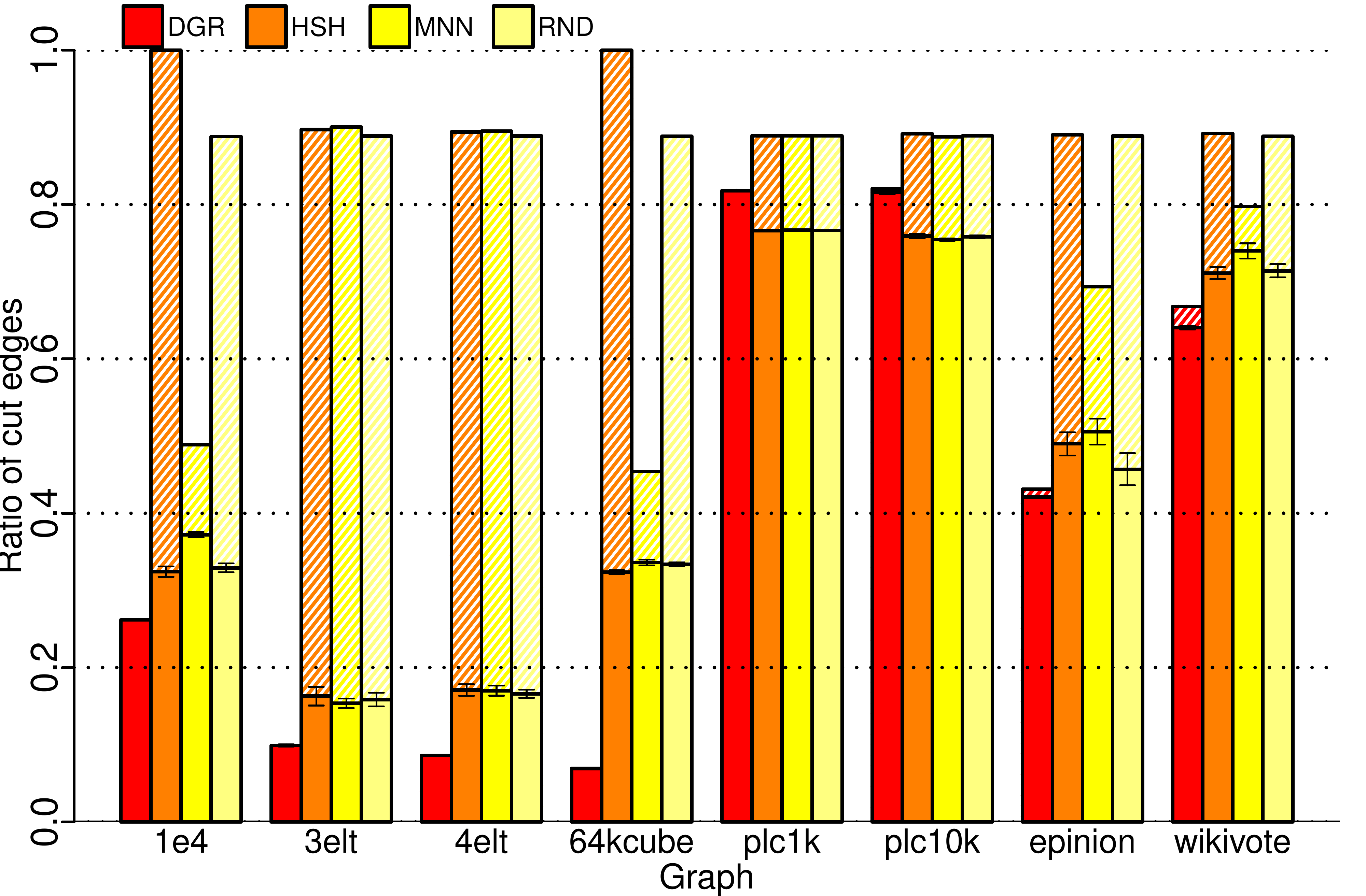}
       \vspace{-10pt}
  \caption{ Ratio of cut edges for each graph after running the iterative heuristic over four different initial partitioning strategies. Clearer bars show the ratio with the initial partitioning}
         \vspace{-10pt}
  \label{fig:41-accumulated}
\end{figure}

Figure \ref{fig:41-accumulated} provides the results obtained for each graph on average of 10 experiments. Each group of four bars shows the results for a different graph, partitioned with the four initial partitioning strategies. The graph overlaps the bars for the cut ratio of both the initial partitioning (dashed colour fill), and the improved final partitioning after running the adaptive heuristic (filled colour). The improvement obtained by the iterative heuristic (if any) from the initial partitioning is represented by the visible dashed colour bar.

Looking at HSH initial partitioning, the iterative heuristic significantly improves the cut ratio for FEM graphs (the five leftmost bar groups), with greater than $0.6$ improvement. Adaptive partitioning also provides substantial improvements for RND and MNN partitioning strategies. When applying it to a state-of-the-art initial partitioning technique (DGR), it only slightly improves the cut ratio, since the heuristics have a very similar (greedy) nature. It is worth noting that for large-scale graphs, DGR and MNN require full graph knowledge, which poses significant limits to its scalability and its applicability to real deployments \cite{Ugander}. For real use cases we have used hash, as it is the de facto standard used by most other large-scale partitioning systems. However, we believe these results show that  the adaptive heuristic is compatible with state-of-the-art partitioning techniques, and would optimise the partitioning when graph dynamism is required (see Figure \ref{fig:2-dyngraphs}). 

Looking now at power law graphs (the four rightmost bar groups), non-DGR initial partitioning is also improved by the iterative heuristic. For these types of graphs the final cut ratio is higher than the levels achieved on FEM graphs. The results show that DGR could not provide either a low cut ratio for these graphs, showing that they are more difficult to partition.

\subsubsection{Convergence Study}

\begin{figure}
  \centering
    \includegraphics[width=0.98\columnwidth]{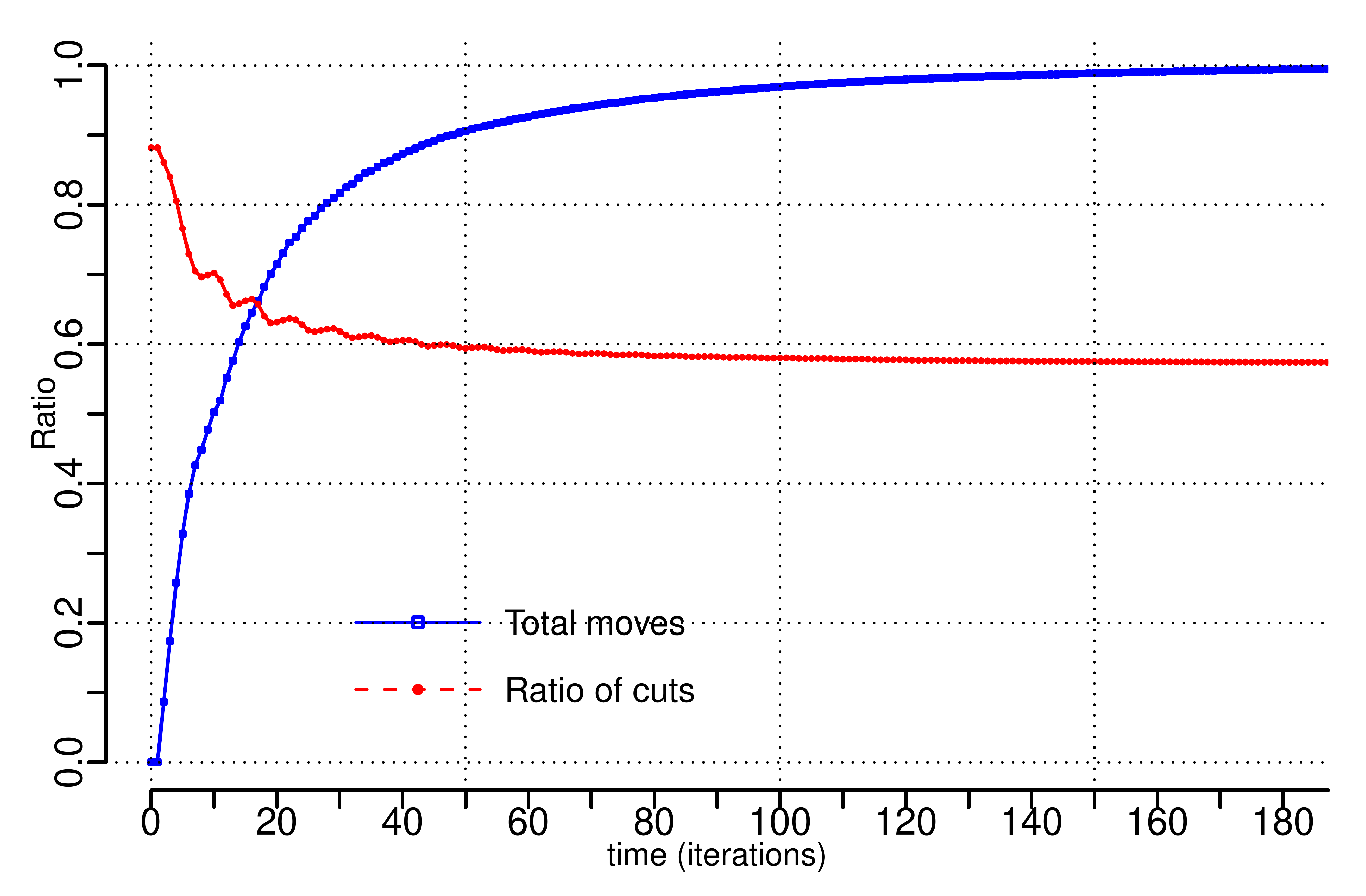}
    \vspace{-10pt}
  \caption{ Accumulated migrations and cut ratio evolution on Livejournal graph. }
 	\vspace{-15pt}
  \label{fig:migrations}
\end{figure}

We study how the adaptive heuristic converges to a partition distribution. To this end, we collected the number of migrations between vertices at each iteration, as well as the evolution of the ratio of cut edges. Figure \ref{fig:migrations} shows cut ratio (red), and ratio of migrations completed (blue) for the Livejournal graph. The graph was partitioned using modulo hash. Cumulative vertex migrations are triggered rapidly in the initial iterations, with more than 50\% of the moves completed at the tenth iteration. During this stage the cut ratio is also improved to less than 0.7 from the initial 0.9. The rate of migrations slows down rapidly, and it takes to iteration 47 for the heuristic to migrate 90\% of the vertices. At this stage, 90\% of the ratio of cuts improvement has been achieved. 

We observed similar behaviour in the improvement of cut ratio and triggered number of migrations with different graphs and partitioning strategies. The first iterations of the heuristic trigger the majority of the movements, as well as a significant part of the improvement in the partitioning. This has an important impact in the system performance. As the cut ratio is reduced, computation performance will improve thanks to the reduced communications cost. However, migrating vertices brings an additional overhead that might cause performance bottlenecks if too many migrations happen at the same iteration. The dampening factor of $s$, migration quotas and the deferred migration technique help to smoothen the initial peak of migrations. 

It can be predicted that from a performance point of view the initial iterations will be affected the most by the additional overhead. In this setting iteration execution times quickly go down as the cut ratio improves and later iterations will quickly improve computation execution performance, as both the overhead from vertex migrations goes down and the quality of partitioning quickly improves, therefore optimising computation execution time. We present the observed relationship between migrations, quality of the partitioning and performance in the following subsection.

\subsubsection{Performance of Adaptation to Dynamic Changes}

\begin{figure}
  \centering
    \includegraphics[width=8.0cm]{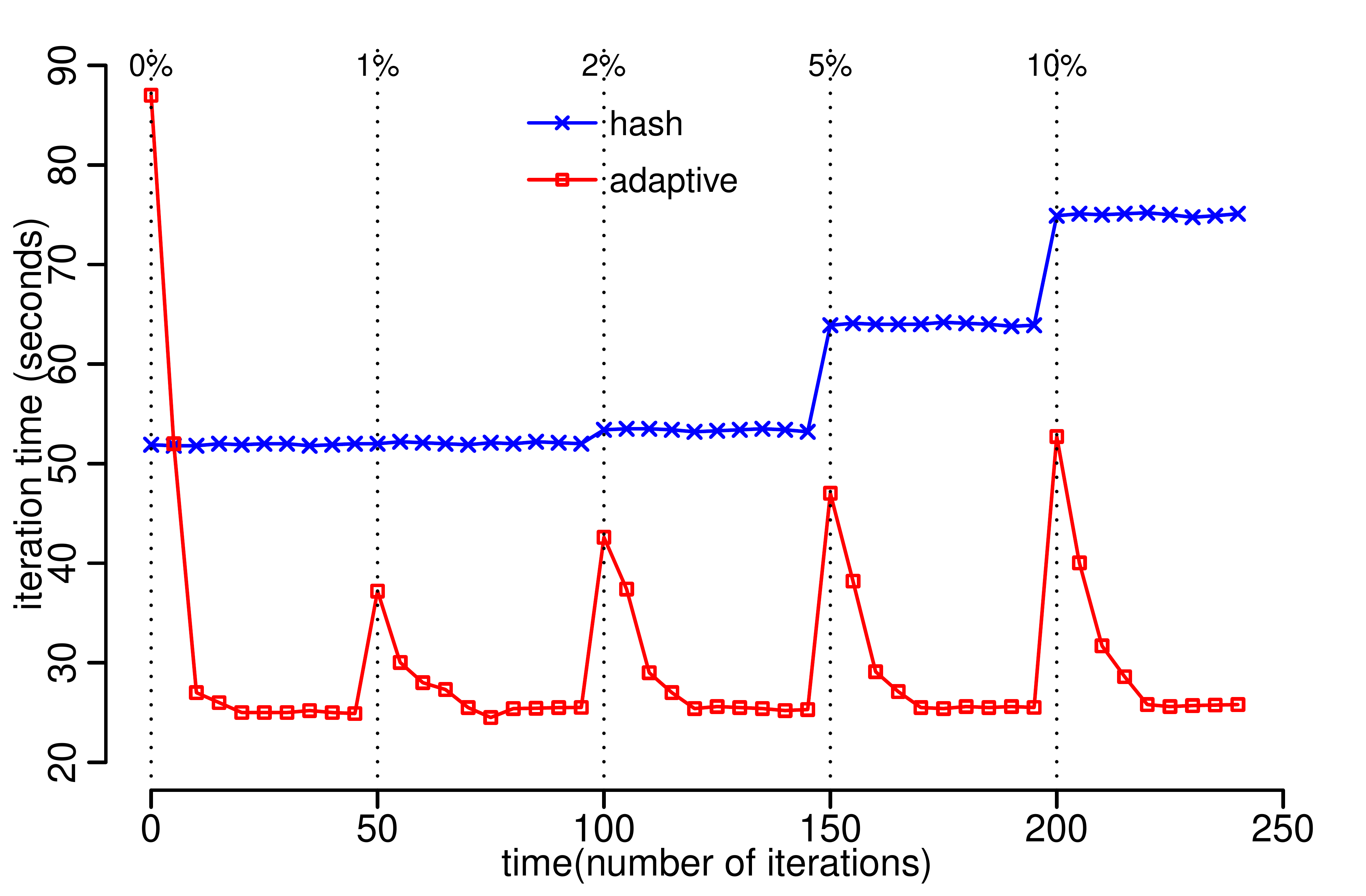}
    \vspace{-5pt}
  \caption{ Execution time evolution after injecting changes to the Livejournal graph. }
 	\vspace{-15pt}
  \label{fig:livejournal}
\end{figure}

We evaluate the performance (in iteration time) of the system can adapt to controlled changes in the graph. We loaded the Livejournal graph to our system, partitioned initially with modulo hash, and calculated the estimated diameter, with the heuristic used in \cite{Mizan}. Every 50 iterations we injected to the running graph a burst of new vertices (based on a forest-fire expansion), each addition increasing the current graph size by 1, 2, 5 and 10\%, respectively. The graph was continuously processed to compute the diameter of the graph. Figure \ref{fig:livejournal} shows the average time per step (in minutes) on LiveJournal with a static approach (hash partitioning - blue) and our dynamic heuristic (red). Time is divided in five regions, where we can observe the adaptation of the heuristic to the initial partitioning, and each one of the changes to the graph. 

At the initial stage (0\%), we can observe the performance impact of the convergence behaviour of the algorithm we just discussed. Over the initial 10 iterations, where 50\% of migrations take place, the overhead from migrations significantly affects computation performance, with the first five running almost 80\% slower than the hash baseline, and the following five at roughly the same time. The next five iterations show a substantial improvement, with the average time going down to 54\% of the time required by the hash baseline. The following iterations show considerably smaller improvements in the iteration execution time. 

We can conclude that the performance overhead from the heuristic is strongly dependent on vertex migrations. The decision part of the heuristic is executed at every iteration, and does not seem to overweights the benefits from an improved partitioning. For performance reasons, it will be worth to adapt the graph when computations will take place for an extended number of iterations. Otherwise, it is possible that the initial overhead will cancel any benefits form achieving a improved partitioning.  

Now let us observe the changes on execution time when we inject graph changes. First, looking at the static partitioning, execution time increases, growing up to an increase over 50\% from the initial time. On the other hand, the adaptive heuristic shows similar behaviour for each graph injection. Initially execution time degrades due to the migrations overhead, but quickly the graph is adapted  and the execution time returns to figures almost identical to the ones obtained with the initial graph (0\%). Larger additions to the graph inflict higher performance degradation over the first subsequent iterations, although after 10 iterations the heuristic has returned to values close to the optimal. The exact nature of the migrations overhead is heavily system dependent, but it becomes more taxing to the system the more abrupt changes occur.

\subsection{Real-Word Use Cases}
\label{subsec:usecases}

We validated our system with a set of real-world use cases. We aimed at testing the scalability of the system, and the capability to cope with dynamic graphs in different scenarios.  We believe that the diversity in the workloads of each application helps to support the general validity of our approach.

In all three cases, we ran the same experiments on two deployments of our system. One with the adaptive partitioning heuristic, and one with static hash partitioning. 

\paragraph{\textbf{Adaptation in Real-Time Social Network Analysis}}

Our first use case evaluates the capability of the system to analyse a dynamic graph modelled after a continuous stream of information. We aim to assess the adaptation capabilities of the heuristic, with respect of the evolution in the quality of the partitioning, and the impact in execution time. 

\begin{figure}
\centering
\includegraphics[width=0.99\columnwidth]{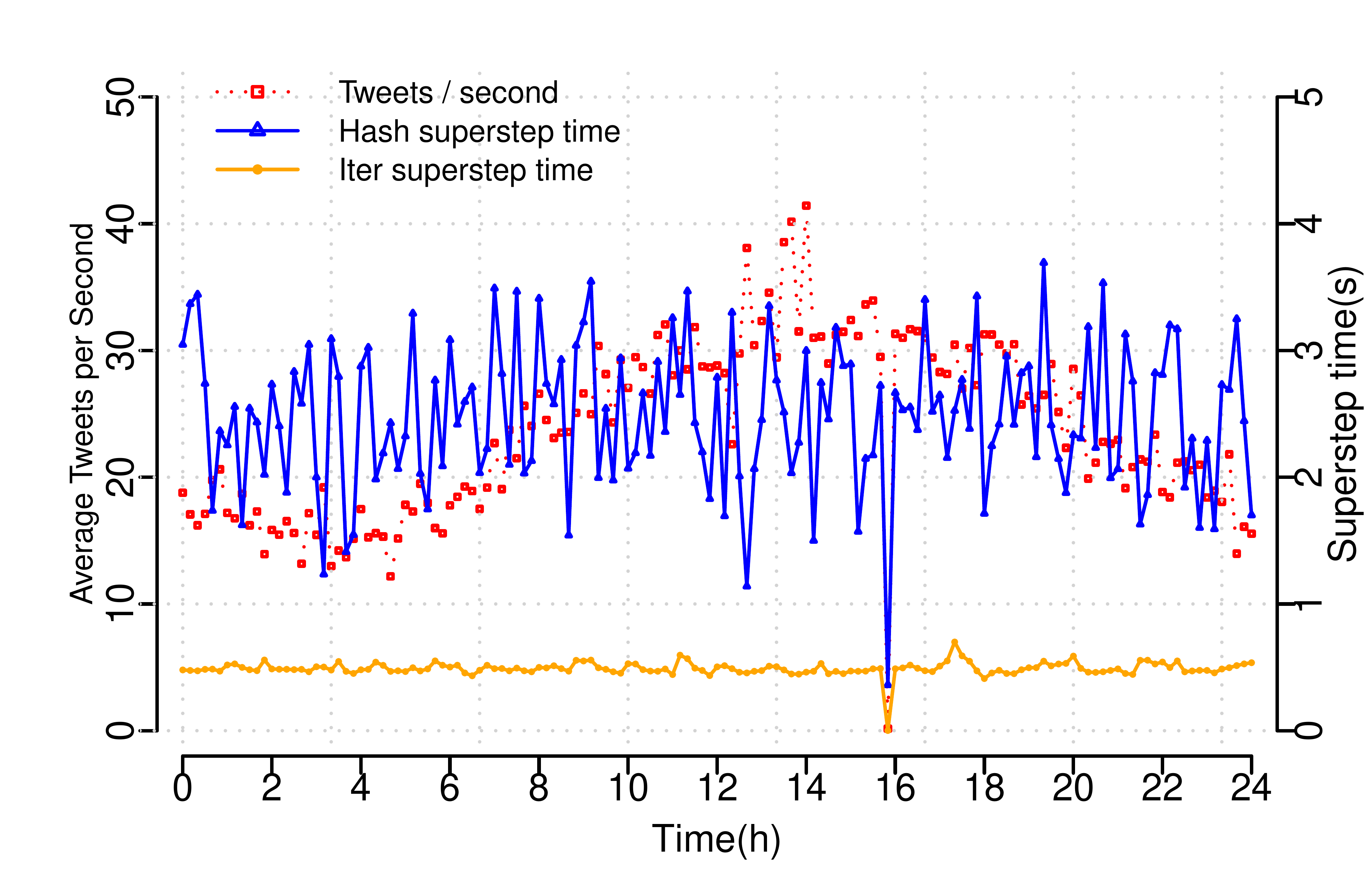}
\vspace{-15pt}
\caption{Throughput and performance obtained by processing the incoming stream of tweets originated from London. Each point represents the average of 10 min of streaming data. }
\vspace{-15pt}
\label{fig:twitter}
\end{figure}

We captured tweets in real-time from Twitter Streaming API, and built a graph where edges are given by mentions of users. Over this power law graph, we continuously estimated the influence of certain users by using the TunkRank heuristic \cite{TunkRank2009}. In this test, execution time is bound by the number of messages sent over the network at any point in time (over 80\% of the iteration time)

We ran the experiment simultaneously in two separate clusters: One cluster used the adaptive heuristic, while the other used static hash partitioning instead. In Figure \ref{fig:twitter} we can observe the average results from processing tweets collected in the London area over a whole day (Friday, 5th Oct 2012), after running continuously for 4 days. The red line shows the rate at which tweets are received and processed by the system, while the blue and orange lines show average execution times per iteration, with and without adaptation, respectively. The sudden drop in throughput and iteration time is due to a failure in one of the workers that led to the triggering of xDGP recovery mechanism.

As can be observed, the average execution time is significantly improved when applying the adaptive heuristic, (mean of 0.5 secs instead of 2.5 secs, including the added overhead). Importantly, the optimisation of the partitioning with only local information significantly lessens variability in execution times, by reducing the impact of network communications (more neighbours are local).

\paragraph{\textbf{Adaptation in Mobile Network Communications}}

The second use case shows how xDGP can support online querying over a large-scale dynamic graph. We used a dataset from a mobile operator, with one month of mobile telephone calls. The dynamic graph was created by applying a sliding window to the incoming stream of calls as follows: Nodes represent users and calls are modelled as edges between these users. Therefore, new calls add nodes and vertices to the graph and both are removed from the graph if they are inactive for more than the window length (one week). The window size yielded weekly addition/deletion rates of 8 and 4\%, respectively, which is higher than those reported in previous studies due to the shorter period of analysis \cite{Cortes03computationalmethods}. 

Over this graph, we continuously computed the maximum cliques of each node. The maximum clique was obtained as follows: In the first iteration, each vertex sends its lists of neighbours to all its neighbours. On the next iteration, given a vertex $i$ and each of its neighbours $j$, $i$ creates $j$ lists containing the neighbours of $j$ that are also neighbours with $i$. Lists containing the same elements reveal a clique. As these lists can get large, this heuristic produces heavy messaging overhead for large graphs, especially if these are dense, and not negligible CPU costs, although not as much as the biomedical use case described later. The main problem of applying the iterative heuristic to this use case is that optimising message passing in iteration 1 places neighbours together and creates hotspots (all the members of a clique will be calculating the same cliques in parallel on the same host). To reduce duplicate calculations (reduce ``hot zones'') only lists for $j>i$ are created and only neighbours of neighbours with ID $j>i$ are added to those lists.
 
In contrast with the previous scenario, this application requires freezing the graph topology until a result is obtained, therefore  buffering all the graph changes until the computation finishes (for two iterations instead of one). Meanwhile, adaptation occurs at every iteration.This characteristic makes the scenario more challenging than the previous one, as every iteration will trigger the adaptation to a batch set of changes to the graph. Call data was streamed into the system with a speed up factor of 15, to increase the amount of buffered changes per cycle , further testing the adaptation capabilities of the heuristic. 

\begin{figure}[t]
\centering
\includegraphics[width=0.99\columnwidth]{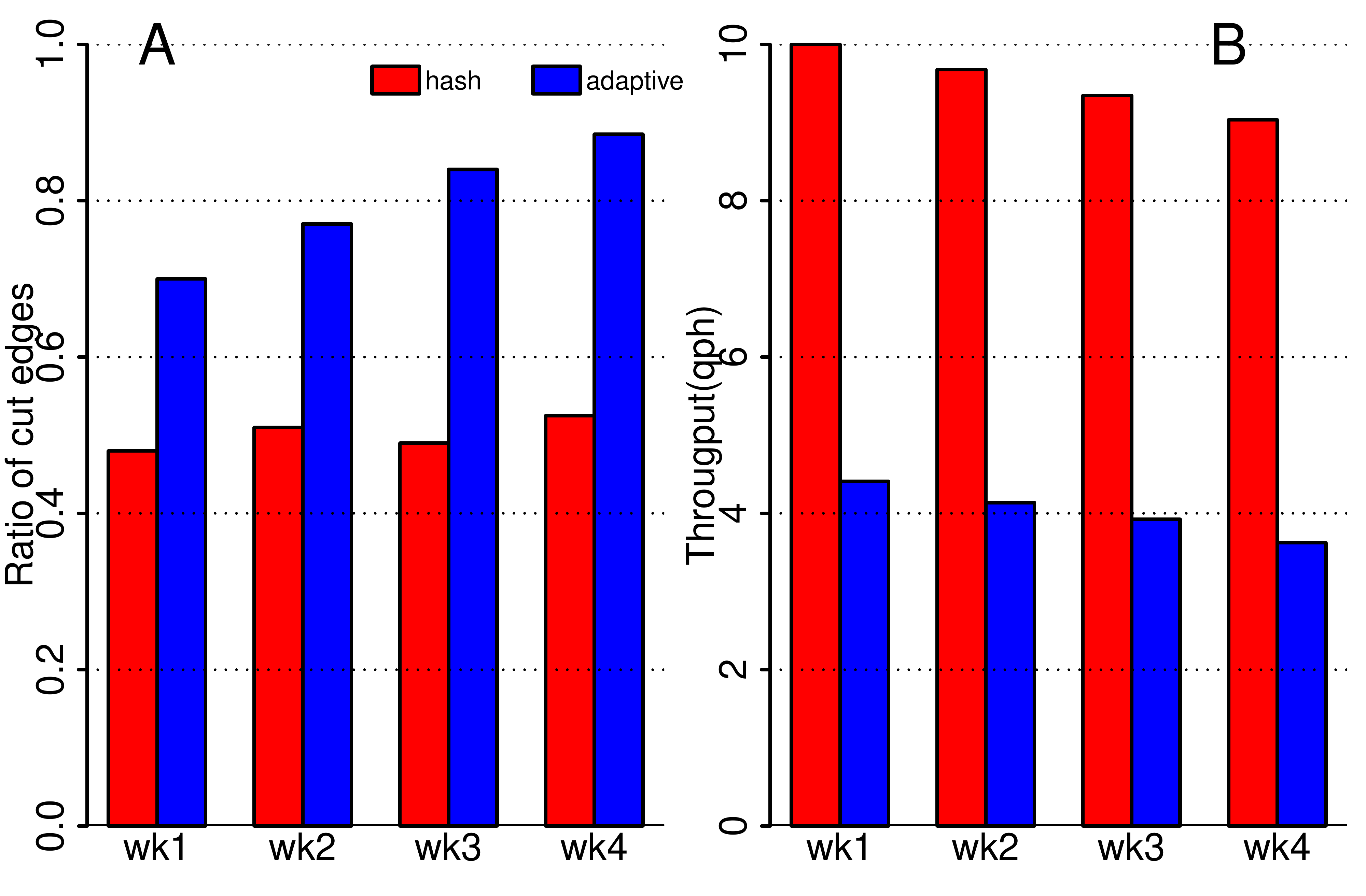}

\caption{Evolution of the ratio of cuts \textbf{(Left)} and average iteration (step) time \textbf{(Right)} during the 4 weeks of available data. The experiments were performed in a cluster of 5 workers (96GB RAM, 10 GbE, 12 cores).}
\vspace{-15pt}
\label{fig:phonePerf}
\end{figure}

We run the clique finding application in two separate clusters, with and without the adaptive heuristic.  Figure \ref{fig:phonePerf} shows the weekly results in both cases. It can be seen that the adaptive partitioning maintained a stable number of cuts, resulting in consistently higher throughput (more than twice the throughput provided by hash portioning). Moreover, weekly trends show that the static scenario experiences further performance degradation over time due to the higher cut ratio.

\paragraph{\textbf{Adaptation in Biomedical Simulations}}

The final scenario assesses the suitability of the proposed system and heuristic for implementing large scale biomedical simulations. Biomedical simulations require long-running computations, with heavy CPU usage. Simulations are often implemented on specialised clusters, using message-passing libraries such as MPI. The use case presents a different type of application (long-running simulations), that operates at a considerably higher scale than the previous scenarios.

The input graph is a 100 million vertex/300 million edges FEM representing the cellular structure of a section of the heart. Each vertex computes more than 32 differential equations on one hundred variables representing the way cardiac cells are excited to produce a synchronised heart contraction and blood pumping \cite{Tusscher2004}. The graph state occupies a total of 3TB in memory among the 63 worker machines running the simulation. Using static hash partitioning (without the adaptive heuristic), simulation time is still dominated by the exchange of messages (more than 80\% of the time), even though CPU time is not negligible (more than 17\%).  The iterative heuristic works in this use case similarly to the other experiments, achieving a final speedup of 2.44 after convergence.

At a certain point in the simulation, we mimic the effect that adding stem cells differentiating into cardiac tissue would have. These new cells are injected to the graph as an additional 10M vertices and 30M edges, joining the tissue in the border of the infarcted region, based on preferential attachment. These changes bring the total memory usage close to full occupation in the cluster. 

\begin{figure}[t]
\centering

   \includegraphics[width=8cm]{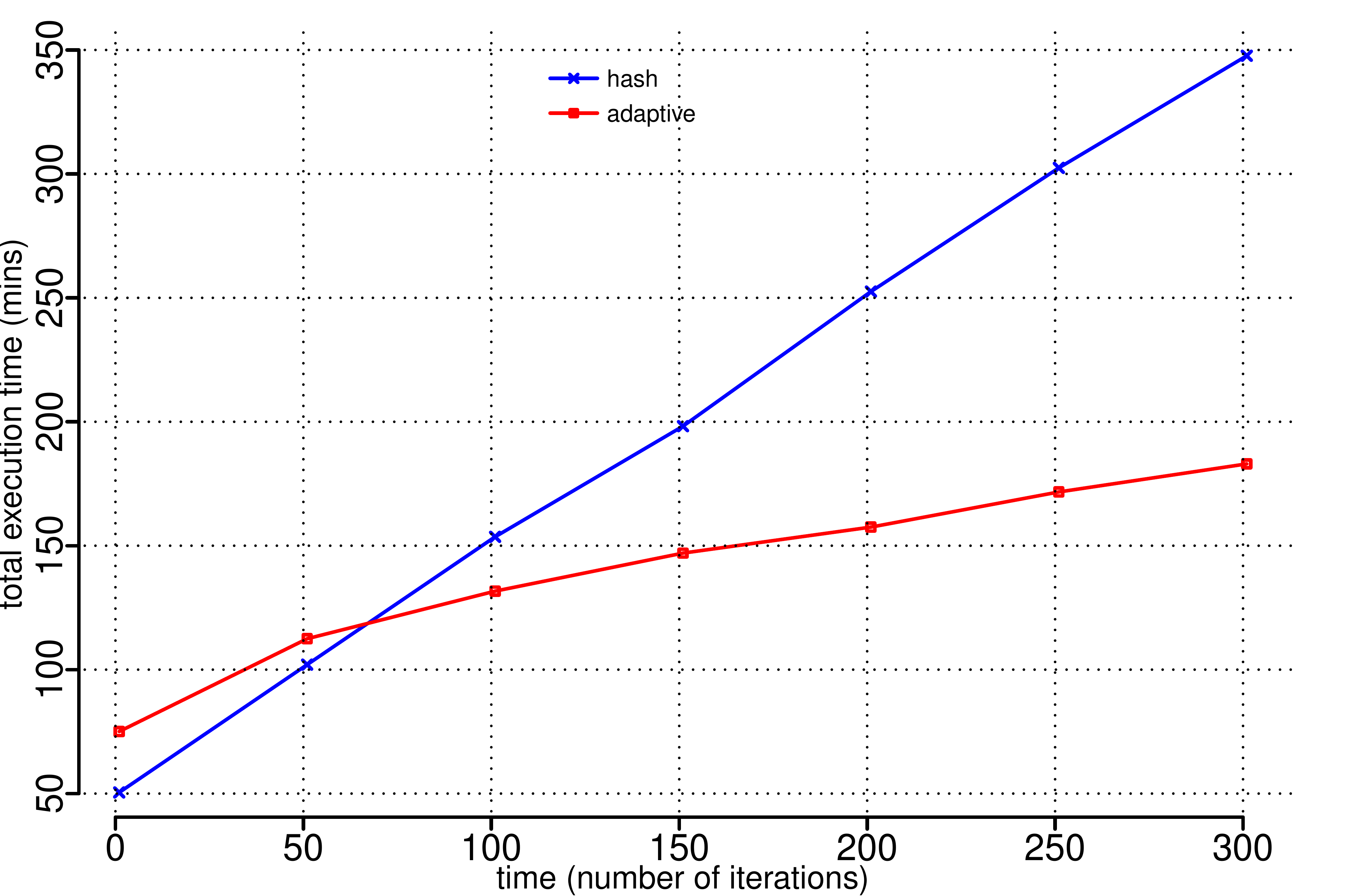}
   \caption{Accumulative execution time after expanding the heart cell FEM using a forest fire model (extra $10^7$ vertices and $3*10^7$ edges. Results were obtained in a cluster of 63 workers (64GB RAM, 10GbE and 12 Cores)}.
\vspace{-20pt}
\label{fig:bio}
\end{figure}

We show in Figure \ref{fig:bio} the cumulative execution time, from the instant changes were added to the graph structure, for both a static hash partitioning (blue) and our iterative heuristic (red). As expected, the first iterations are affected by the overhead from the triggered vertex migrations, but in the long term the improved partitioning significantly shortens simulation time. Comparing these results with previous use cases the heuristic performs better on continuously changing graphs. It will be worth to adapt to abrupt changes in the graphs only when facing long-running computations, such as biomedical simulations.

\section{Related Work}
\label{sec:related}

To the best of our knowledge, there is no system that continuously processes large-scale dynamic graphs, while adapting the internal partitioning to the changes in graph topology. Our system adapts to large-scale graph changes by repartitioning while 1) greatly reducing the number of cut edges to avoid communication overhead, 2) producing balanced partitioning with capacity capping for load balancing, and 3) relying only on decentralised coordination based on a local vertex-centric view. The heuristic is generic and can be applied to a variety of workloads and application scenarios. 

\subsection{General Partitioning}

The idea of partitioning the graph to minimise network communication is not new and it has inspired several techniques to co-locate neighbouring vertices in the same host \cite{Bhatt1983, Pellegrini1996, Karypis1998,Newman06,Blondel2008,Arora2009,DasSarma2010}. These approaches try to exploit the locality present in the graphs, whether due to the vertices being geographically close in social networks, close molecules establishing chemical bonds, or web pages related by topic or domain, by placing neighbouring vertices in the same partition. The parallel version of METIS \cite{metis}, ParMETIS \cite{parmetis}, leverages parallel processing for partitioning the graph, through multilevel k-way partitioning, adaptive re-partitioning, and parallel multi-constrained partitioning schemes, but requires a global view of the graph that  greatly reduces its scalability \cite{Salihoglu2012}. Other techniques have been explored that study graph properties projected onto a small subset of vertices \cite{Leskovec2007,DasSarma2010}. These may be effective in some particular contexts, but they are not broadly applicable. 

\subsection{Changing Graph Structure at Runtime}

Beyond these fixed techniques for static graphs, the need to continuously adapt to changes to the graph structure without the overhead of re-loading an updated snapshot of the graph or re-partitioning from scratch, has been recently reported in practical \cite{rostoker2007,Mucha2011,gaci2012} and more theoretical \cite{Nicosia2012} studies. Few systems can cope with run time changes in structure \cite{Najork2009,Sedge,Cheng2012}. However, these cannot handle structural graph changes, either degrading partition quality or fully triggering the partitioning process.  

\subsection{Initial Partitioning Strategies}

Some techniques try to alleviate performance degradation by optimising partitioning during the initial loading of the graph in memory (i.e. they do not adapt in run time). For instance, in \cite{Stanton2012} authors evaluate a set of simple heuristics based on the idea of exploiting locality, and apply them on a single streaming pass over a graph, with competitive results and low computation cost. The authors show the benefits of this approach in real systems. In addition to adaptations to changes in structure, some systems dynamically adapt the partitioning of the graph to the bandwidth characteristics of the underlying computer network to maximise throughput \cite{Chen2012}. Mizan \cite{Mizan} ignores the graph topology and instead optimises the system by performing runtime monitoring and load balancing. The graph processing system finds hotspots in specific workers and migrates vertices  to a paired worker who have the highest number of outgoing messages in an attempt to balance the load. 

GPS \cite{Salihoglu2012} applies the technique most similar to ours from the heuristic point of view, but system implementation limits its application to static graphs. There are two main differences: 1) they allow vertices to move while an iteration is still running, while we move vertices between two consecutive steps; 2) to simplify location of a migrated vertex, they modify the ID. This prevents adding new elements, since their ID may conflict with one of a previously loaded and migrated vertex. We preserve de ID of the vertex by using a more complex vertex localisation mechanism, which enables near real-time changes in the topology and subsequent optimisations to increase vertex locality.

Initial partitioning strategies only optimise the starting graph, with several of these techniques requiring some extent of global information. This poses significant scalability problems \cite{Ugander}, whereas our approach relies only on local information. Additionally, as shown in Figure \ref{fig:2-dyngraphs}, these approaches cannot cope with changes that alter the structure of the elements already loaded (e.g. node and vertex deletion) or keeping optimal partitioning as the graph changes.  



\subsection{Dynamic Adaptation beyond Initial Partitioning}

In addition to adapting the initial partitioning of the graph, some systems attempt to keep a small overhead when processing changing graph structures. In  \cite{Ugander},  partitioning was optimised in slowly changing graphs, with changes being applied the next time the graph was loaded. The authors employ a label propagation mechanism enhanced with geographical information to improve graph partitioning. The process involves linear programming, being computationally very expensive (reported calculations of 100 CPU days) and implying global aggregation of local (vertex-level) utility functions.

Sedge \cite{Sedge} is a dynamic replication mechanism (as opposed to a re-partitioning one). Sedge keeps a fixed set of non overlapping partitions and then dynamically creates new ones or replicates some of them in different machines to cope with variations in workload. Replicated systems are more focused at providing low latency to multiple concurrent and short-lived queries. Our system tries to keep a few long-lasting (continuous) queries which results are modified as a consequence of changes in the information or the topology of the graph.

\section{Conclusions}
\label{sec:conclusions}

Real world graphs are dynamic, and mining information from graphs without considering the evolution of their structure over time can have a significant impact on system performance.

In this work we have focussed on adapting to graph changes in a highly scalable way, while working under the challenges of migrating vertices in a distributed system. The presented heuristic adapts the graph partitioning to graph dynamics at the same time as computations take place. We show through our experiments that the heuristic improves computation performance (with higher than 50\% reduction in iteration execution time), adapting to both continuous and abrupt changes. 


A key performance factor for adapting to graph changes is the tradeoff between the additional overhead incurred by repartitioning the graph, and the effective performance improvement from a better graph partitioning. We have found vertex migration to be the predominant source of overhead (specially when migrating a high number of vertices), and we will work on further system optimisations for efficient vertex creation and migration. 

We believe this work contributes a first step towards the study of the performance of systems based on dynamic graphs, but there is significant work ahead. The space of dynamic graphs is still not well known, neither from models that characterise the structure and temporal dimensions of their growth, nor from deeply characterising the performance implications, finding better techniques to decide when and how to adapt to changes. 

\section{Acknowledgments}

The authors would like to thank  \'{A}lvaro Navas and Steve Uhlig for their continuous support and advice during the elaboration of this work.

\bibliographystyle{abbrv}
\bibliography{socc13}

\end{document}